# Universal amorphous-amorphous transition in $Ge_xSe_{1-x}$ glasses under pressure

Can Yildirim[1,2], Matthieu Micoulaut[2], Punit Boolchand[3], Innokenty Kantor[4,5], Olivier Mathon[4], Jean-Pierre Gaspard[1], Tetsuo Irifune[6], and Jean-Yves Raty[1]

[1] SPIN - Université de Liège, Institut de Physique B5, 4000 Sart-Tilman, Belgium
[2] Laboratoire de Physique Théorique de la Matière Condensée, Paris Sorbonne Universit´es, UPMC, 4, place Jussieu 75252 Paris Cedex 05, France
[3] School of Electronics and Computing Systems, College of Engineering and Applied Science, University of Cincinnati Cincinnati, OH 45221-0030, USA
[4] European Synchrotron Radiation Facility, 6 rue Jules Horowitz, BP 220, 38043, Grenoble Cedex, France
[5] Technical University of Denmark, Fysikvej 307, 2800 Kgs. Lyngby, Denmark and
[6] Geodynamics Research Center, Ehime University, Matsuyama 790-8577, Japan

**Abstract:** Pressure induced structural modifications in vitreous $Ge_xSe_{100-x}$ (where 10≤x≤25) are investigated using X-ray absorption spectroscopy (XAS) along with supplementary X-ray diffraction (XRD) experiments and ab initio molecular dynamics (AIMD) simulations. Universal changes in distances and angle distributions are observed when scaled to reduced densities. All compositions are observed to remain amorphous under pressure values up to 42 GPa. The Ge-Se interatomic distances extracted from XAS data show a two-step response to the applied pressure; a gradual decrease followed by an increase at around 15-20 GPa, depending on the composition. This increase is attributed to the metallization event that can be traced with the red shift in Ge K edge energy which is also identified by the principal peak position of the structure factor. The densification mechanisms are studied in details by means of AIMD simulations and compared to the experimental results. The evolution of bond angle distributions, interatomic distances and coordination numbers are examined and lead to similar pressure-induced structural changes for any composition.

**INTRODUCTION**

Pressure-induced phase transformations have attracted widespread interest in condensed matter science. Analogous to the polymorphism observed in crystalline materials, the existence of metastable configurations in an amorphous system is called polyamorphism [1]. The amorphous-amorphous transformations (AATs) have been detected in a variety of systems including oxide and chalcogenide glasses [2-8] amorphous Si [9] and Ge [10] and even water [11]. AATs have been often observed in systems having directional bonding and low coordination numbers, such as tetrahedrally bonded systems that allow for a large free volume in the system, more recently this phenomenon was also identified in a binary metallic glass [12].

Among network-forming glasses, vitreous germanium chalcogenides are of special importance due to their technological applications. In contrast to to their oxide counterparts, a considerable amount of homopolar bonds exist in the structure of these glasses, enhancing the structural variability [13]. In particular, the binary Ge-Se system, which could be considered as an archetype for covalently bonded networks [14] with a wide glass forming range (43 Ge %) that allows for tuning topological properties linked with short and intermediate range atomic ordering provided by the stiffening of the network structure [15]. The main building blocks of the network in Ge-Se glasses are $GeSe_{4/2}$ tetrahedra, which are connected either by edges or corners such as in $GeSe_2$ [6,16,17]. This particular composition has been studied under pressure by means of X-ray and neutron diffraction, Raman scattering as well as acoustic measurements [6,7,18,19]. A conversion from edge sharing (ES) to corner sharing (CS) tetrahedra has been found along with a continuous increase in the coordination number up to 10 GPa. Recently, an EXAFS study has shown a semiconductor to metal transition between 10-15 GPa accompanied with a profound increase in the coordination number and Ge-Se distances [20]. Similar observations have been reported for $GeS_2$ under pressure with earlier EXAFS experiments [21] and atomic simulations [22].

It has been also shown that during compression, unlike the continuous coordination increase observed in the stoichiometric compound $GeSe_2$, the coordination number remains constant up to 8.6 GPa [23]. A more recent study suggests a transition from a low density amorphous semiconductor to a high density metallic amorphous between 10 and 15 GPa using X-ray diffraction (XRD) and electrical conductivity measurements [8]. Moreover, additional evidence for a semiconductor to metal transition in $GeSe_4$ under pressure was noted recently using a combination of neutron diffraction and ab initio simulations [24]. That work confirmed that the average coordination number shows no significant changes up to $\approx 8$ GPa and that non-negligible fractions 5-fold Ge and 3-fold Se appear around 13 GPa.

Despite the experimental findings and simulations carried out regarding the pressure-induced structural changes in $GeSe_2$ and $GeSe_4$ glasses, a complete understanding of the underlying densification mechanisms leading to the AATs in $Ge_xSe_{100-x}$ glasses is still lacking. Here we report results of a comprehensive study of the polyamorphic transitions in vitreous $Ge_xSe_{100-x}$ with a detailed description of the structural modifications using a combination of X-ray absorption spectroscopy (XAS), XRD and ab initio molecular dynamics (AIMD) simulations under pressure up to 41.4 GPa. Our results provide a thorough insight of the atomic packing rearrangements triggered during the AAT over a range of compositions including flexible and optimally rigid networks.

**RESULTS**

**X-ray diffraction.** XRD results for $Ge_{25}Se_{75}$ with increasing pressure as an example for all compositions are shown in Fig. 1 a). A strong diamond diffraction peak due to DAC is observed at q=3.04 Å$^{-1}$, which was subtracted through the empty cell measurement from the measured XRD spectra of the samples in order to find the peak positions reliably. The q range of the diffractograms allows one to distinguish the typical features of $Ge_xSe_{100-x}$ glasses; namely the first sharp diffraction peak (FSDP) located at around 1.0 Å$^{-1}$ which has been attributed to Ge-Ge correlations [25] and the first principal peak (PP) at ≈2.0 Å$^{-1}$. With increasing pressure, the FSDP loses its intensity and vanishes at around 4 GPa due to the collapse of the network structure whereas the PP notably gains intensity, sharpens and shifts to higher q values. It has been argued that the intensity of the principal peak in glasses can be attributed to the development of the extended range ordering during densification controlled by Ge-Se and Se-Se correlations [26]. No sign of crystallization was observed up to a maximum pressure of 41.4 GPa for $Ge_{17}Se_{83}$ and up to 20 GPa for the other compositions studied. The PP positions that have been found through Gaussian peak fits as a function of increasing pressure are given in the Fig. 1 b). It can be seen that at ambient pressure the position of the PP increases as the network stiffens (i.e. increasing Ge content). A more detailed analysis shows that two different trends exist in the densification of $Ge_{17}Se_{83}$ which could be traced with the linear lines. A rather fast increase of the PP position with a slope of 0.0267±0.0015 Å$^{-1}$ GPa$^{-1}$ and a slow down having a slope of 0.0098±0.0008 Å$^{-1}$GPa$^{-1}$ after ≈15 GPa. This change in slope indicates the existence of a pressure induced transition in structural modifications.

**EXAFS.** X-ray absorption near edge structure (XANES) spectra at Ge and Se K edge for vitreous $Ge_xSe_{100-x}$ at ambient pressure are shown in Fig. 2 a) and 2 b), respectively. At Ge K edge, a prominent absorption peak implying the transition between 1s orbital and the unoccupied states (the white line) to be located at ≈11.107 keV with a shoulder peak found at ≈11.113 keV. Similarly at Se K edge, a strong absorption peak is located at 12.657 keV followed by a shoulder at 12.665 keV. Fig. 2 c) and d) exhibit the EXAFS spectra of $Ge_{17}Se_{83}$ glass at different selected pressures at Ge K and Se K edges, respectively. The changes in the oscillations with increasing pressure can be seen clearly.

The evolution of the average Ge-Se interatomic distance for $Ge_{17}Se_{83}$ and $Ge_{20}Se_{80}$ as a function of increasing pressure extracted from the back Fourier EXAFS data is compared with the AIMD results for $Ge_{18}Se_{82}$ in Fig. 3 a). For $Ge_{17}Se_{83}$ at pressures below 13 GPa the average Ge-Se bond length shortens of about 0.02 Å. This bond compression is followed by an abrupt increase in Ge-Se bond length that is observed in the pressure range 15 GPa ≤ P ≤ 20 GPa resulting in an elongation of about 0.09 Å. Similarly for $Ge_{20}Se_{80}$ a bond compression is followed by a sudden increase in the Ge-Se distance at around 17 GPa and continues to increase up to ≈20 GPa, reaching a value of ≈2.44 Å. This result lies in line with the previous study on $Ge_{20}Se_{80}$ [8]. In contrast to $Ge_{17}Se_{83}$, due to lack of pressure points at P≥20 GPa, the decrease of Ge-Se bond length in the high density amorphous (HDA) phase is not observed. Nevertheless, further increase in the applied pressure should lead to a bond compression in the high-pressure structural motifs as shown for $Ge_{17}Se_{83}$. At this stage, in the light of XRD and EXAFS results under pressure, evidence of a threshold behavior in the structural response of $Ge_{17}Se_{83}$ to the applied pressure is detected around 17 GPa both in reciprocal and real space. The AIMD results of $Ge_{17}Se_{83}$ agree well with the experimental under pressure, providing a very similar trend. Note that the Ge-Se interatomic distance at ambient pressure is underestimated compared to the values extracted from EXAFS. These minor differences that occur due to the local dependence of the exchange correlation kernel on the electronic density are generally accepted in AIMD simulations as noted in earlier studies [27,28]. The average Ge-Se interatomic distance decreases under pressure values up to ≈10 GPa. Subsequent to this decrease, a dramatic bond elongation is observed upon further compression and around 15 GPa, the average Ge-Se bond length springs back to its ambient pressure value. This increment is followed by a similar rate of change until ≈22 GPa and eventually a drop in the slope of the curve is observed towards 32 GPa. Invariably, other flexible compositions that are generated by AIMD have shown a similar behavior under pressure ; an initial bond compression in the low density amorphous (LDA) phase in the early stages of densification followed by an evident transition monitored by an increase in the average Ge-Se bond length.

Fig. 3 b) reveals the pressure dependence of the energy variation in Ge K edge position of $Ge_{17}Se_{83}$. A gradual shift in the edge position towards lower energies of about s-1.4 eV is followed by a limited decrease at P ≥ 20 GPa. The negative shift in the edge position is attributed to the semiconductor to metal transition which was shown in chalcogenides under pressure with EXAFS [20,21] and electrical conductivity [8] measurements. In the inset of Fig. 3 b) the clear change of the red shift is demonstrated at selected pressures. Another feature observed in Ge K edge XANES for the compositions under consideration is the area decrease in the shoulder peak of the white line. On the other hand, the shoulder peak in the Se K edge XANES spectra is observed to shift to higher energies (i.e. increasing about ≈6.5 eV up to 19.3 GPa for $Ge_{20}Se_{80}$) which may be an indication of the changes in the inter-layer Se-Se correlations.

**Equation of State.** We have analyzed the structural changes in terms of AIMD simulations in order to track the alterations in the atomic arrangements responding to the applied pressure. In Fig. 4 we provide the calculated volume change of glassy $Ge_xSe_{100-x}$ AIMD trajectories as a function of pressure and compare them with experimental measurements where available. The results show that the equation of state of simulated amorphous $Ge_xSe_{100-x}$ are very well reproduced. The simulation data were fitted by both second and third-order isothermal Birch Murnaghan (BM) equation of state (EoS) ; for $Ge_{20}Se_{80}$ the fits yields to a bulk modulus $B_0$ of 10.51±0.31 GPa and 11.04±0.76 GPa with the first pressure derivative $B_1$ being 4.25±0.15, respectively. These results agree well with the experimental findings which was shown to be $B_0$ = 10.4±0.14 GPa and $B_1$ = 6.0±0.11 for $Ge_{20}Se_{80}$ [23]. As for $Ge_{25}Se_{75}$, a second-order BM fit gives a $B_0$ of 10.64±0.18 GPa. Note that these values were obtained by fitting to all pressure points. It has been shown earlier that LDA phase has a smaller bulk modulus compared to that of high density amorphous (HDA) phase [8]. In this respect a value of 9.44±0.02 is obtained for $Ge_{20}Se_{80}$ when a second-order BM EoS is fitted in a range of 0 to 9.7 GPa, where the LDA to HDA transition is detected. Similarly a LDA bulk modulus is calculated as 11.09±0.23 GPa for $Ge_{18}Se_{82}$ which is in good agreement with the result obtained by a direct volumetric study for a similar composition $Ge_{17}Se_{83}$ yielding a bulk modulus of 11 GPa [29].

**Pair Distribution Functions.** The computed real space properties of $Ge_{25}Se_{75}$ are presented in Fig 5. The evolution of coordination numbers with increasing pressure scaled to reduced densities $\rho/\rho_0$ with $\rho_0$ the density at ambient pressure, is shown in Fig 5 a). The coordination numbers at ambient pressure calculated by AIMD satisfies the 8-N rule (N being the number of outer shell electrons); Ge having 4 and Se having 2 neighbors in the first shell. However, a distinguishable Se coordination change takes place at around $\rho/\rho_0 \approx 1.4$, accompanied by the increase of Ge coordination at $\rho/\rho_0 \approx 1.6$. These changes are connected to the transition of the low-pressure to high-pressure modification. During this stage Ge atoms reach an octahedral coordination (with a considerable fraction of five-fold Ge) while the Se coordination develops and majority of the atoms become 3-fold coordinated (again with some fraction of 4-fold Se). At about 22 GPa the average coordination of Ge and Se atoms reach to 4.87 and 2.59, respectively.

The calculated pressure dependence of the partial pair distribution functions ($g_{ij}$ where i,j = Ge,Se) of $Ge_{25}Se_{75}$ are shown in the other panels of Fig. 5. For the case of $g_{GeGe}$, three main peaks are noticed at ambient pressure at 2.46 Å, 3.02 Å and 3.61 Å, respectively, the first one being associated with homopolar Ge-Ge bonds. Note that these are absent at ambient pressure for lower Ge content. The second peak of $g_{GeGe}$ is assigned to the edge sharing (ES) tetrahedra distances whereas the third peak is linked with tetrahedral corner sharing (CS) bonding distances.

During compression, the first peak associated with homopolar Ge-Ge bonding distance remains nearly the same below 3 GPa and notably increases until reaching a maximum at around 9 GPa. Up to this point the homopolar Ge-Ge bonding has an overall shortening of about 0.06 Å. Further densification causes a decrease in the intensity whereas the peak position starts to increase at 10.5 GPa and reaches back to ≈2.42 at around 22 GPa. Unlike the first peak, the ES peak loses intensity upon compression while shifting to shorter distances and at 4 GPa it completely vanishes. For the case of the CS tetrahedra distances, again a two stage behaviour is observed ; a shift to shorter distances in the peak position until 10.5 GPa followed by an increase upon further compression.

The main feature of the $g_{GeSe}$ function is the strong peak located at 2.36 Å which was also shown to be found at the same distance for the stoichiometric compound $GeSe_2$ by previous neutron diffraction [30] and EXAFS [20] studies, and defines the Ge-Se correlation distances within a tetrahedron. Upon compression significant changes take place in the peak shape and position. In particular, the intensity of the peak slightly decreases at around 3 GPa. This decrease is followed by a substantial intensity loss at 4 GPa, followed by a gradual intensity loss until 8.9 GPa. Another significant drop is observed at this stage, corresponding to the rapid coordination increase and the peak broadens evidently. Of special importance is the evolution of the peak position of the first peak of $g_{GeSe}$ with increasing pressure. A bond compression similar to what is observed in the Ge-Se pairs in $Ge_{17}Se_{83}$ extracted from EXAFS signals is present below 8.9 GPa. Up to this point the Ge-Se interatomic distance decreases of about 0.04 Å. Subsequent to this drop, a marked increase in the Ge-Se distance can be traced accompanying the rapid coordination increase indicating the high pressure modification transition. At the final pressure point attained, Ge-Se distance reaches to a value of ≈2.40 Å. Finally $g_{SeSe}$ exhibits two main contributions at the ambient pressure. The first peak is located at 2.37 Å defining the homopolar Se-Se bond whereas the second is found at 3.85 Å that corresponds to the distances associated with the edge of the tetrahedra. The bond length again starts to increase drastically after 8.9 GPa in a manner similar to the pressure behavior of the Ge-Se bond length. On the other hand the second peak shifts to shorter distances under compression. The distance associated with edges decreases gradually reaching to a minimum value of 3.40 Å at 21.8 GPa. The detailed analysis on the atomic trajectories indicates that this distance can be identified as the edges of an octahedron.

The partial pair distribution functions of the other compositions respond in a similar way to the applied pressure. In particular, AIMD results showed that the ES peak of the $g_{GeSe}$ vanishes for all compositions and a "homopolar" Ge-Ge peak appears while the CS peak shifts to shorter distances. Moreover, the average Ge-Se bond length evolution extracted for $Ge_{20}Se_{80}$ and $Ge_{10}Se_{90}$ lies in line with Fig. 5 a) : a gradual bond compression followed by a significant increase around ≈15 GPa. To get further insight on the resemblances of the densification mechanisms, we now turn to detailed investigation of the neighbour distance variations under compression. Fig. 6a shows the computed evolution of the average distance decomposed into the neighbors around a Ge atom of selected composition as a function of reduced density. As the applied pressure increases the bond length of the second shell neighbors, the fifth and the sixth neighbors, decrease for all compositions. However, the first shell neighbors display smaller variations, and an interesting feature is that the distance evolution of the second shell neighbors starts to decrease more rapidly around $\rho/\rho_0 = 1.4$, where they effectively become the part of the first shell. Consequently, these results show that the transition from the low pressure to the high pressure modification is triggered by the merging of the second shell neighbors which is followed by a substantial increase in the coordination number at a certain threshold. This mechanism applies to all the compositions under consideration.

**Bond Angle Distributions.** Fig. 6 also shows the variations in the bond angle distributions (BAD) of $Ge_{20}Se_{80}$ under pressure. Here $Ge_{20}Se_{80}$ illustrates the common behavior observed for all compositions generated by AIMD. Fig. 6b shows that at ambient pressure, two main peaks are present in the Ge-Se-Ge BAD located at ≈80° and 100° that are associated with ES and CS tetrahedra, respectively. During compression, the former loses intensity in contrast to the increment in that of the latter. At around 9.6 GPa the peaks merge and further compression leads to a broad peak centred at $90^0$ with a shoulder at 100° as the system globally transforms from a tetrahedral to an octahedral coordination. A closer look into the atomic trajectories indicates that the shoulder peak at 100° is associated with the Ge-Se-Ge angles in the rings that contain five atoms. Fig. 6d exhibits the Se-Ge-Se BAD under pressure. A strong peak centred at ≈110° defining the tetrahedra. The intensity of this peak reduces during the low pressure modifications. An obvious dramatic change occurs between 6.7 GPa and 9.6 GPa, and the main peak shifts to ≈100°, this change being connected to the marked increase of the coordination number. Further compression yields to the continuation of the shift towards smaller angles reaching to the octahedral angle, 90º (confirmed by the emergence of a smaller peak around 170-180 degrees). Fig 6c further analyses the changes by decomposing the BAD into neighbor contributions of the form i0j where (i,j) represent labelled neighbors. The average BAD of the first shell neighbors, 102, 103 and 104 contributions, that are around ≈110º at ambient pressure remains nearly at the same angle with slight fluctuations until $\rho/\rho_0 \approx$ 1.4. Again, further densification induces a coalescence in the BAD of the first and the second shell neighbor contributions, reaching approximatively 90º. At $\rho/\rho_0$ > 1.4, the increment in the BAD of the 105 and 106 contributions are somewhat similar.

**Coordination Numbers.** The tetrahedral to octahedral conversion can be traced by looking at the coordination geometry around a Ge atom. Fig. 7 reveals the evolution towards octahedral order as a function of reduced density for $Ge_xSe_{100-x}$. The inset of the figure exhibits the population between 4, 5 and 6 folded Ge atoms in $Ge_{18}Se_{82}$ under compression. The tetrahedral configuration at ambient pressure is maintained at a constant value for all the compositions under consideration during the low-pressure modifications. Indeed, the percentage of 4 fold coordinated Ge atoms is maximum for the isostatic composition $Ge_{20}Se_{80}$. The small variations from a perfect tetrahedral connection observed in the other compositions are due to 2- or 3- folded atoms. Upon compression, around ≈ $\rho/\rho_0$= 1.2 an increase in the amount of 5 fold coordinated Ge atoms is detected, while the Ge-Se bond compression still takes place. The onset of the LDA to HDA phase transformation is found to be located at $\rho/\rho_0$=1.4 for all compositions, manifested by the occurrence of the octahedra. The simultaneous decrease in the number of tetrahedral order accompanying the appearance of 6 fold coordinated Ge atoms can be observed in the inset of Fig. 7. A fully octahedral coordination is not achieved within the compression range of our simulations, reaching up to 32 GPa for $Ge_{18}Se_{82}$. The other important feature that can be deduced from of Fig 7 is the composition dependence of the rate of change of tetrahedral to octahedra transformation. The rate of increase of octahedral motifs decreases as Ge content increases. In other words, the AAT becomes more diffuse as the network stiffens.

Finally, we show the AIMD computed average coordination number normalized to the value at ambient pressure (eg. $\acute{n}$ = 2.2 for $Ge_{10}Se_{90}$ at $\rho/\rho_0$=1) as a function of reduced density in Fig. 8. Again, the trend in the average coordination numbers is the same for all compositions studied; a marked increase in the coordination after $\rho/\rho_0$ >1.4, signifying the LDA to HDA transformation. A transition zone is suggested in Fig. 8 below which only the low-pressure modifications take place. This result demonstrates a universal behavior in the densification mechanisms for target $Ge_xSe_{100-x}$, and indicates the transition from a low density amorphous to high density amorphous around $\rho/\rho_0 > 1.4$ regardless of the Ge content.

**DISCUSSION**

Our XRD measurements show no sign of crystallization under pressure, thus the amorphous nature of the structure is preserved up to 20 GPa for all compositions, in particular up to 41.4 GPa for $Ge_{17}Se_{83}$. However, an increase in the intensity of the PP upon increasing pressure is observed. The intensity of the PP has been related to extended range ordering in previous studies [26]. In addition to this, the vanishing of the FSDP in the early stages of low-pressure modifications indicates the collapse of the IRO, as reported earlier [17]. A significant observation out of our XRD analyses is the shift in the PP position under pressure. We have detected a two-step behavior appearing around 15 GPa for $Ge_{17}Se_{83}$ agreeing well with what was observed for $Ge_{20}Se_{80}$ in a previous study [8]. Note that in order to distinguish the two-step slope in the PP position, one should have sufficient number of pressure points in the HDA region.

The EXAFS data gave us access to crucial structural information such as the evolution of the average Ge-Se interatomic distance under pressure. The bond compression in LDA phase is followed by a sudden increase in the Ge-Se bond length. The pressure at which this jump occurs corresponds to the slope change observed in the PP position (Fig. 1b), suggesting existence of another densification mechanism taking place in at another lengthscale. The Ge-Se bond elongation can be linked to the tetrahedral to octahedral transformation, and support to this argument is revealed from the AIMD results. Therefore, a clear evidence for a transition from a LDA to a HDA phase occurring within an interval of 15-20 GPa is observed. Moreover, within the compression range of our simulations, a considerable amount of 5 folded Ge was found in the network. From the EXAFS data, a bond compression in the HDA phase is identified with the average Ge-Se bond length shortening after 20 GPa for $Ge_{17}Se_{83}$. Even though this bond shortening is not reproduced in the pressure range for the AIMD of $Ge_{18}Se_{82}$, the observed slowing down in the increment of the average Ge-Se bond length with increasing pressure coincides with the decrease in the amount of 5 folded Ge atoms. Thus the degree of octahedral order continues to increase during the compression of the HDA phase along with a simultaneous decrease of 4 and 5 folded Ge atoms after a certain threshold found to be at $\rho/\rho_0 \approx 1.7$ (inset of Fig. 7).

We also provide information as to how the LDA to HDA transformation kinetics are affected with Ge content. As distinct from the first order phase transformations observed in crystals, the AATs can have jump-like (analogous to first order transformations) or rather sluggish kinetics of transformation depending on the the stress level between the high pressure nuclei and the parent phase. Simple amorphous systems such as a-Si and a-Ge, as well as amorphous water have been shown (ref [31] and references therein) to exhibit sharp AATs whereas network forming oxide and chalcogenide glasses such as $GeO_2$ and $GeSe_2$ have been identified with the wide range AATs under pressure [22]. Our AIMD simulations demonstrate similar outcomes, and the rate of AAT which is governed by the tetrahedral to octahedral transformation is found to depend on the Ge content.

The pressure modifications that take place during the AAT mainly involve the convergence of the second shell neighbors with the first shell neighbors, as shown by our atomic simulations, as also suggested from the partial bond angle distributions (Fig. 6). Furthermore, these structural modifications are observed to take place at a certain threshold when densities are scaled to reduced ones, at $\rho/\rho_0 \approx 1.4$. Consequently, these changes are reflected on average the coordination numbers, and a universal behavior is detected in the average coordination number during the transition from LDA to HDA phase in vitreous $Ge_xSe_{100-x}$ (Fig. 8). Although detailed thermodynamic studies are required to elucidate the physical origin of the threshold $\rho/\rho_0 \approx 1.4$, at the present stage we can only speculate that the coordination changes leading to redistribution of the stress that has been accumulated during the bond compression in the LDA phase ultimately leads to an octahedral configuration which corresponds to a different basin in the energy landscape.

**CONCLUSIONS**

In this work, we have investigated the pressure-induced polyamorphism in vitreous $Ge_xSe_{100-x}$ using a combination of EXAFS, XANES, XRD experiments along with ab initio simulations. We have identified a LDA to HDA phase transition with the elongation in the average Ge-Se bond length using both EXAFS and AIMD simulations. Our results show that the convergence of the second shell neighbors to the first shell gives rise to the marked coordination change can be seen as signature of the AAT. The main outcome of our work is the universal threshold parameter at which the LDA phase starts to give a high density octahedrally coordinated phase, and the transformation is found to be taking place in a range of pressure values. Although our work contributes to the understanding of the polyamorphism in many aspects, further thermodynamic evidences are required to have a full grasp of the underlying physics.

**MATERIALS AND METHODS**

**Sample preparation.** Amorphous $Ge_xSe_{100-x}$ samples (where x=10, 17, 20, 22 and 25) were synthesized using 99.999% Ge and 99.999 % Se pieces mixed in the desired ratio and were sealed in evacuated quartz tubes. The quartz ampules were held vertically in a programmable box furnace for 192 h at 950°C to ensure homogenization. The ampules were water quenched from 50°C above the liquidus temperature. FT Raman profiling was performed on the as-quenched melts to confirm structural homogeneity [32]. The as-quenched melts were then heated in the box furnace at Tg+20°C for 10 min and brought down to room temperature at a cooling rate of 3°C/min. The produced amorphous $Ge_xSe_{100-x}$ samples were kept in a glove box for further manipulation.

**X-ray absorption and diffraction.** The chemical composition and homogeneity of the samples was verified within the limitations of energy dispersive X-ray spectroscopy (EDS) analyses using a field emission scanning electron microscope (FESEM) prior to the X-ray absorption (XAS) and X-ray diffraction (XRD) experiments. XAS and XRD experiments have been performed at the European Synchrotron Radiation Facility at beamline BM23 [33]. High pressure conditions up to 41.4 GPa were achieved by a gas driven membrane diamond anvil cell (DAC) having conical Boehler-Almax design diamond anvils with 300 μm culets. In order to avoid XAS spectra spoiling from sharp diamond anvil Bragg reflections, we used anvils machined from a sintered polycrystalline diamond [34]. Fine pieces of glasses were loaded into rhenium gaskets having 120 μm diameter hole and about 35 μm thickness. Paraffin oil was used as pressure transmitting medium in order to achieve hydrostatic condition. The applied pressure was determined through the ruby florescence method from ruby spheres placed close to the border of the hole. XAS data were collected at Ge K edge (11.1 keV) and Se K edge (12.66 keV) using ion chambers. The X-ray beam spot size focused on the sample was about 5 x 5 μm$^2$.

**Data analysis.** XAS data were analyzed using the ATHENA and ARTEMIS programs of the IFFEFIT package [18,35]. The EXAFS, $\chi(k)$, signal for Ge K edge has been obtained after removing the background modelled by a cubic spline and then normalizing the magnitude of the oscillations to the edge jump. The $\chi(k)$ signals were fitted in k space in an interval of 3.96 to 15.2 Å$^{-1}$. The first shell contribution was isolated by selecting the a back transforming range between 1≤R≤3 Å. A Hanning window function was used for the Fourier transformations. The fitting parameters amplitude reduction factor, $S_0^2$, and energy mismatch, $E_0$, were refined with initial predictions between 0.86 to 1 and 5 to 7, respectively. For XAS refinement of the short range structure, two methods were compared to obtain the best-fitting parameters ; in one method the first shell neighbor coordination numbers were fixed to $n_{GeSe}$

= 4 and $n_{SeGe}$ = 2 whereas in the other the coordination numbers found through ab initio calculations of $Ge_xSe_{100-x}$ glasses under pressure are used. The comparison revealed no significant differences in $\sigma^2$ values and Ge-Se distances (i.e. for the structural parameters in the refining of $Ge_{25}Se_{75}$, when $n_{GeSe}$ is set to 4, the calculated best fit yields $R_{GeSe}$ = 2.366(4) Å and $\sigma^2_{GeSe}$ = 0.00429(3), which is identical to the case when nGeSe = 3.87 determined by the ab initio simulations with a 3 % difference in $\sigma_0^2$ ). Therefore, the former approach has been adopted in which the coordination numbers are fixed to $n_{GeSe}$ = 4 and $n_{SeGe}$ = 2 for the XAS data refinement. The EXAFS data that was extracted have been analysed in terms of a single Ge-Se shell due to the fact that all the compositions under consideration are Ge deficient and the amount of Ge-Ge homopolar bonds are much less compared to the dominating Ge-Se pairs. In order to improve the best fit conditions a two-subshell fits representing Ge-Se and Ge-Ge pairs have been interpreted in the extracted EXAFS data. However this approach did not enhance the goodness of fits compared to single shell fitting therefore we conclude that one can trace the changes in the distances of Ge-Se pair to investigate the structural modifications of $Ge_xSe_{100-x}$ glasses under pressure.

XRD patterns were recorded simultaneously during the XAS experiments at each pressure point with a MAR165 CCD detector having 79 x 79 µm$^2$ pixel size placed at about 164 mm from the sample. A Si (111) monochromator was used to select the wavelength of 0.688 Å spanning a q vector range up to 5.68 Å$^{-1}$. The typical exposure time was set to 600s. An empty cell measurement was performed in order to subtract the background scattering from the diamond and the air. The sample-to-detector distance and the detector tilt angles were determined by using diffraction data from a $Ce_2O_3$ standard. FIT2D [36] software package was used for XRD data reduction.

**Ab initio simulations.** The Car-Parrinello molecular dynamics [37] (CPMD) code was used for the ab initio simulations of $Ge_xSe_{100-x}$ in NVT ensemble for N=250 atoms with corresponding number of Ge and Se species. The atoms were positioned in periodically repeated cubic cells of sizes that recover the experimental glass density (i.e.19.97 Å [32] for $Ge_{10}Se_{90}$). Density functional theory (DFT) was used to describe the electronic structure that evolved self-consistently with time. We have adopted a generalized gradient approach (GGA) using the exchange energy obtained by Becke [38] and the correlation energy according to Lee, Yang and Parr (LYP) [39]. The BLYP approach was used due to its account on valence electron localization effects. The valence electrons have been treated explicitly in conjunction with norm conserving pseudopotentials of the Trouiller-Martins type to account for core-valence interactions. The wave functions have been expanded at the G point of the supercell on a plane-wave basis set having an energy cut-off 20 Ry. A fictitious electron mass of 200

a.u. was used in the first-principles molecular dynamics (FPMD) approach. The time step for integrating the equations of motion was set to 0.1 fs. The temperature control was achieved for both ionic and electronic degrees of freedom using Nosé-Hoover thermostats. The initial coordinates of the atoms have been constructed using the atomic positions in a GeSe crystal. To achieve the correct compositions, randomly chosen Ge atoms were substituted by Se atoms depending on the target composition. Each composition was first melted at T = 2000 K for a period of 22 ps to lose the memory of initial configuration, then simulated along selected isotherms (1373 K, 1050 K, 800 K, and 600 K for 22-30 ps each) before thermalizing at 300 K for 40 ps. Three independent quenches with 1050 K being the starting temperature were performed in order to have a statistical average of the 300 K trajectories and the energetically best structure among these quenches were used to apply compression at 300 K by means of decreasing the cell size. The first 4 ps at each isotherm have been discarded.

**ACKNOWLEDGMENTS**

The European Synchrotron Radiation Facility (ESRF) is acknowledged for providing the beam-time for X-ray absorption and X-ray diffraction experiments through the project HC-1862. C.Y. acknowledges IDS FunMat for the financial support for his PhD project (Project 2013-05-EM) and CEA Grenoble and Pierre Noé for EDS analysis as the industrial partner of the project. The present research benefited from computational resources made available on the Tier-1 supercomputer of the Fédération Wallonie-Bruxelles, infrastructure funded by the Walloon Region under the grant agreement n°1117545. Consortium des Equipements de Calcul Intensif (CECI, funded by F.N.R.S) is acknowledged for supercomputing access. C.Y. acknowledges B. Mantisi for stimulating discussions.

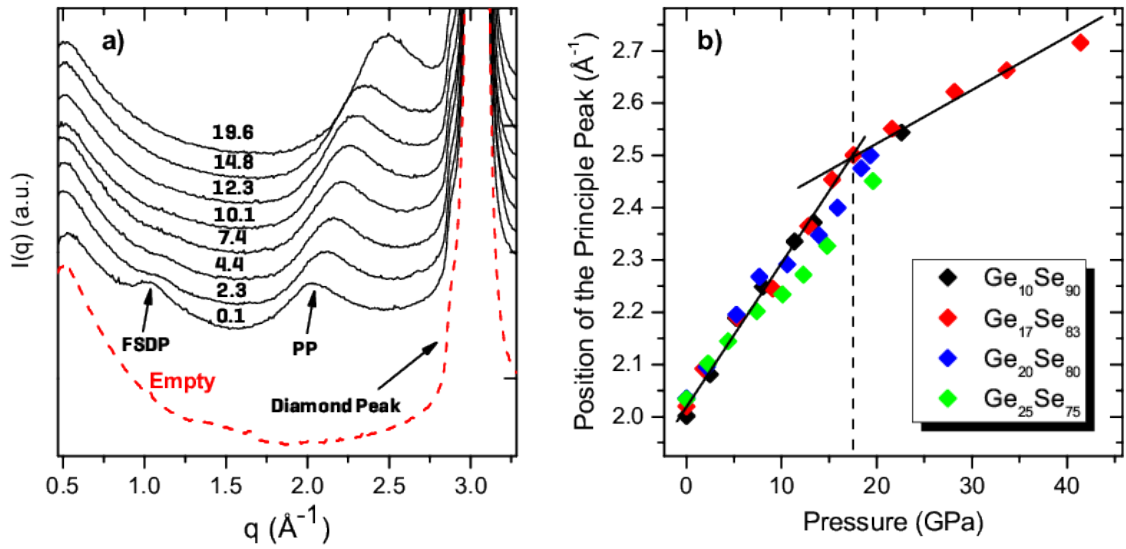

Figure 1: a) X-ray diffraction patterns of $Ge_{25}Se_{75}$ with increasing pressure (written on the continuous curves in GPa unit) and of the measured empty cell (red dashed line). b) The pressure dependence of the principal peak position of the XRD patterns for $Ge_xSe_{100-x}$. The black solid lines are guides to the eye, marking the high and low pressure regimes while the dashed lines indicate the transition in $Ge_{17}Se_{83}$ glass.

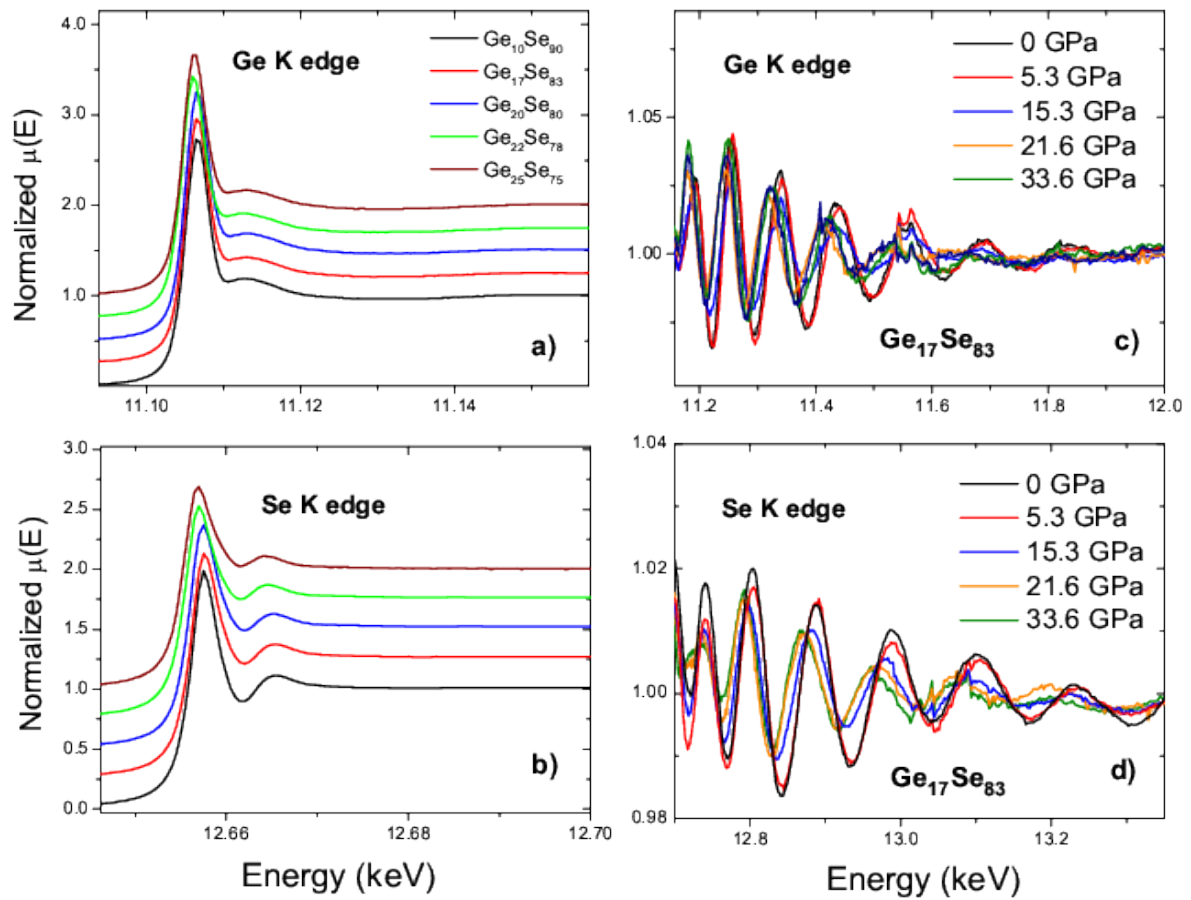

Figure 2: Normalized X-ray absorption spectra at the a) Ge K edge and b) at the Se K edge of amorphous $Ge_xSe_{100-x}$ at ambient pressure. EXAFS oscillations of $Ge_{17}Se_{83}$ at the c) Ge K edge and d) at the Se K edge under selected pressure points collected using DAC.

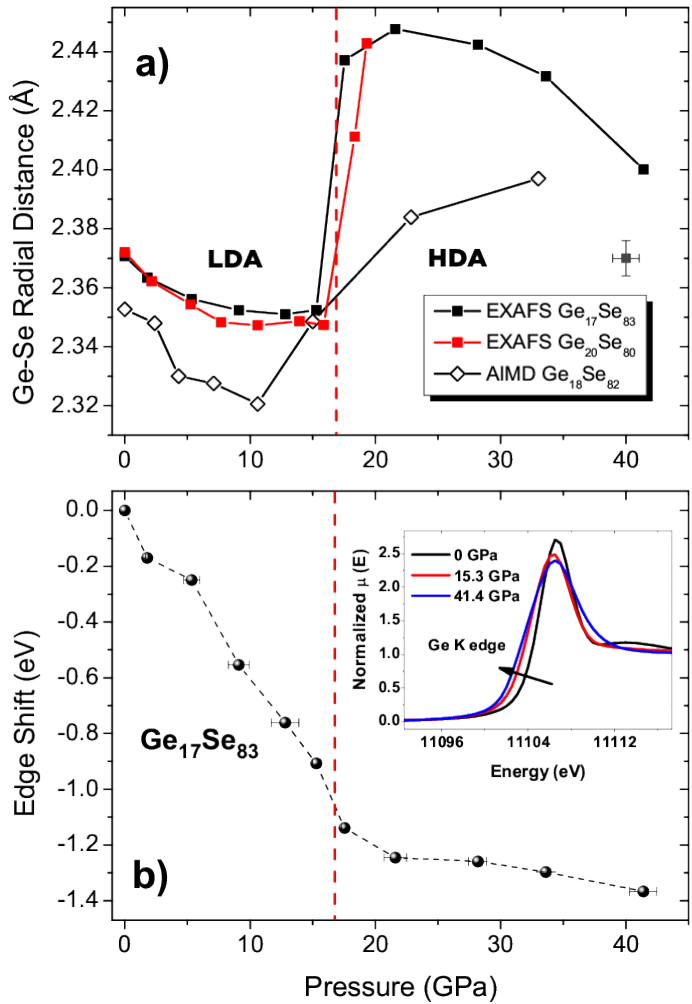

Figure 3: a) Evolution of the Ge-Se interatomic distance as a function of increasing pressure. The filled symbols represent $Ge_{17}Se_{83}$ (black) and $Ge_{20}Se_{80}$ (red). The open diamond symbols are AIMD results for $Ge_{18}Se_{82}$. The estimated error bars are shown separately in the graph. The red dashed line corresponds to the interception of low and high pressure regimes shown by the pressure dependence of the PP position in Fig. 1 b). b) Energy decrease in the Ge K absorption edge position with increasing pressure of $Ge_{17}Se_{83}$. In the inset, the XANES spectra of selected pressures are compared to illustrate the edge shift.

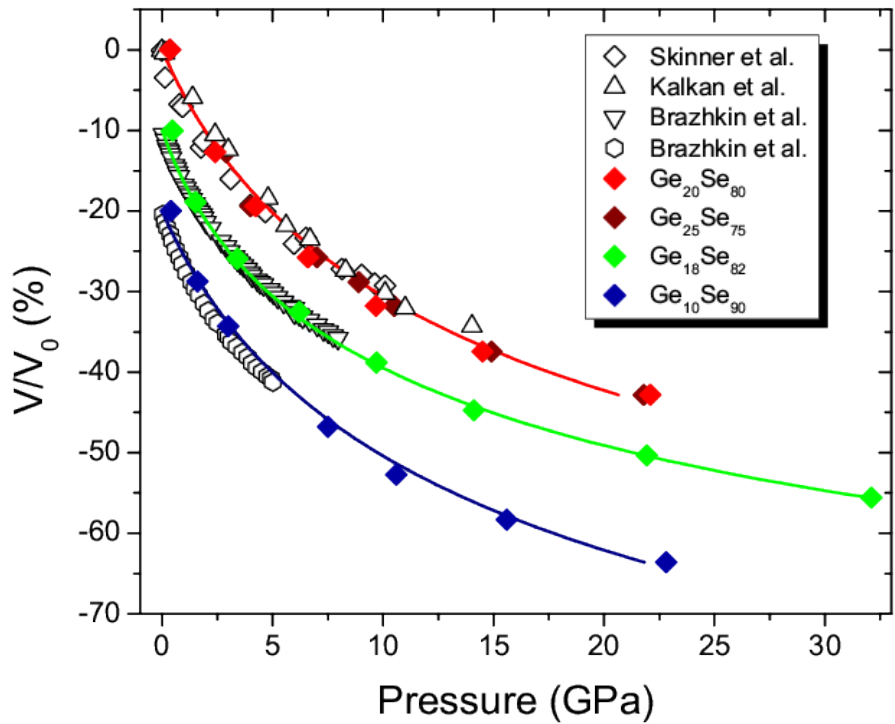

Figure 4: The pressure-volume equation of state for amorphous $Ge_xSe_{100-x}$. The results obtained by AIMD (filled diamond symbols) are compared to the measured data of $Ge_{20}Se_{80}$ of Kalkan et al. [8] and Skinner et al. [23], $Ge_{17}Se_{83}$ and $Ge_8Se_{92}$ of Brazhkin et al. [29] (open symbols) The solid lines represent third-order Birch-Murnaghan equation of state fits to the simulation data and the corresponding data are presented with 10% internal displacement in the reduced volume axis.

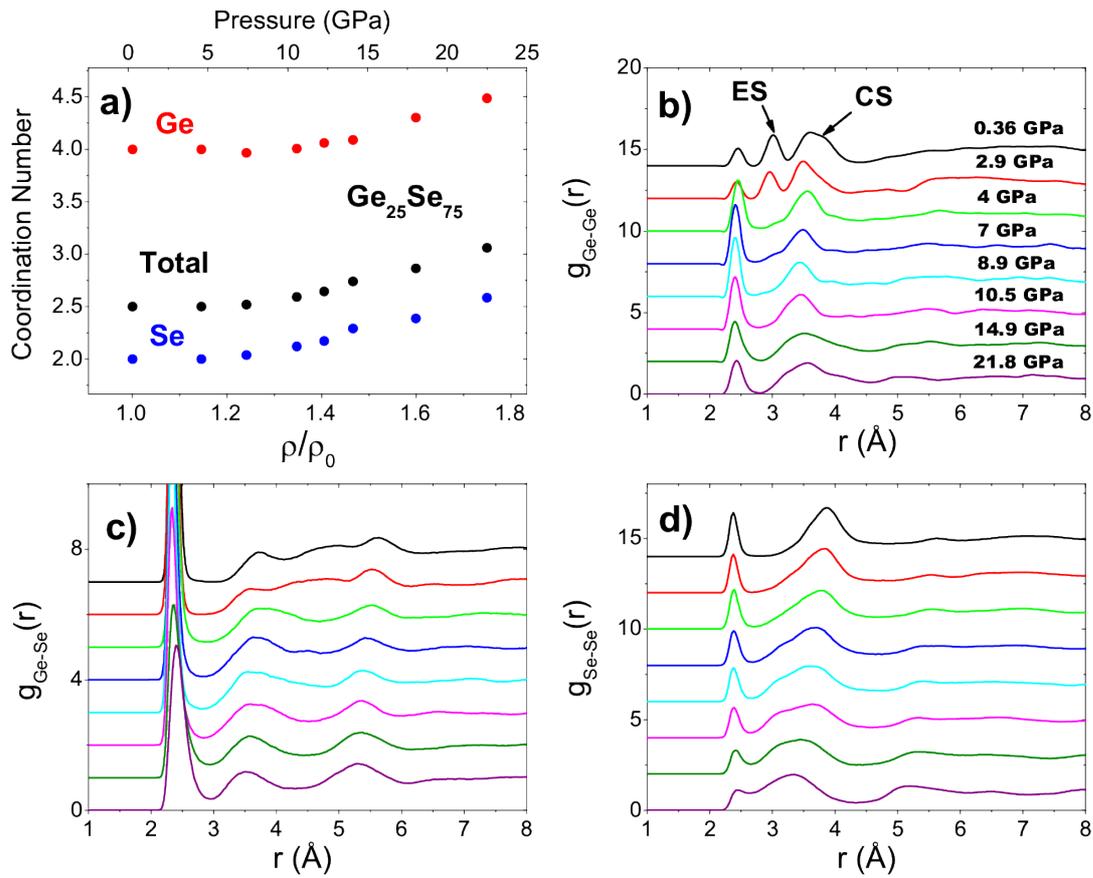

Figure 5: AIMD results of $Ge_{25}Se_{75}$. a) The reduced density dependence of the coordination number. Here, the cut-off distances for the calculation of the coordination numbers are defined by the position of the first minimum of the total pair distribution functions, r = 2.85 Å. The gradual increase in the coordination of Ge and Se atoms becomes pronounced around $\rho/\rho_0$ = 1.4. The calculated partial pair distribution functions b), c) and d) as a function of pressure. Note that the evolution of Ge-Se bond length could also be traced in a trend similar to what is observed in Fig. 4

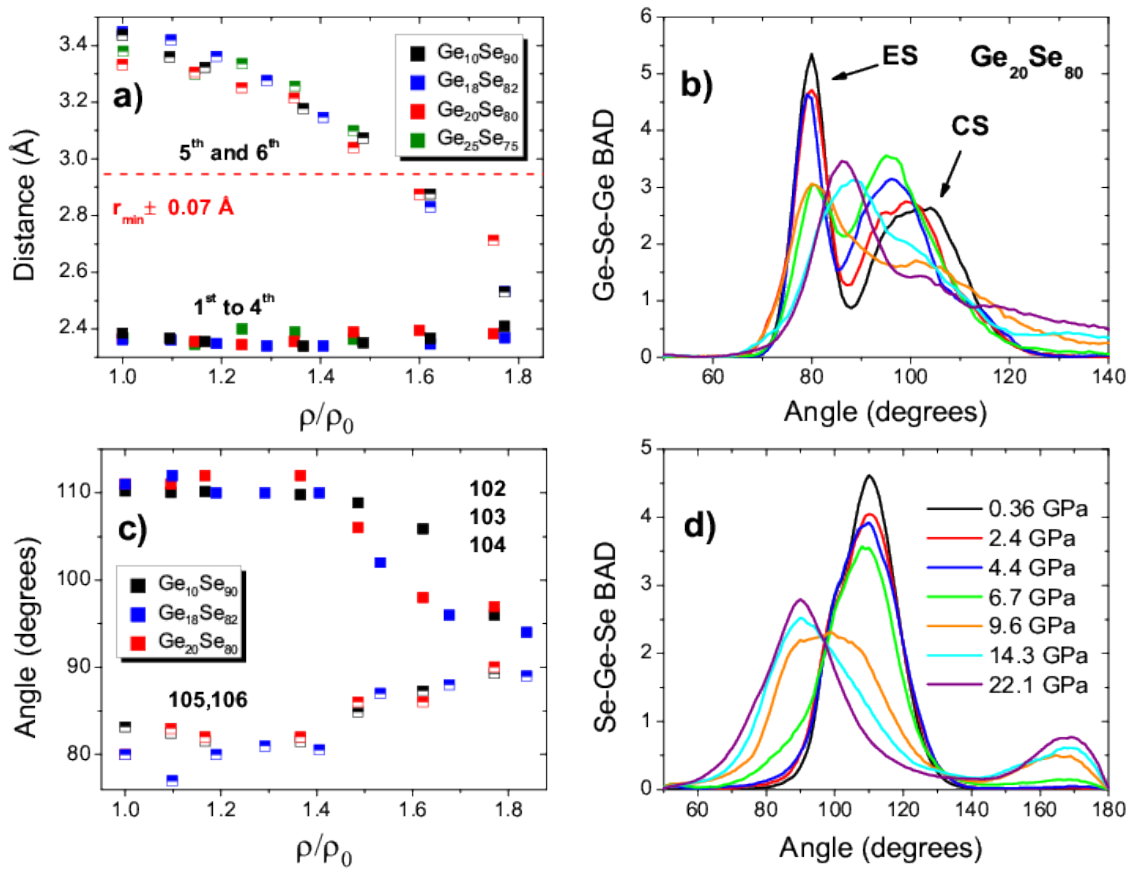

Figure 6: a) Calculated interatomic distances evolution of selected compositions as a function of $\rho/\rho_0$ the filled squares shows the average distances within the first shell consisting of 4 neighbors while the half filled circles show the average distances of the 5th and the 6th neighbors. The dashed red line corresponds to the minimum $r_{min}$ of the pair correlation function, and serves to split the first from the second shell of neighbors. b) Computed bond angle distributions Ge-Se-Ge with increasing pressure for $Ge_{20}Se_{80}$ c) Reduced density dependence of the angle distributions split into neighbor number of the selected compositions obtained from CPMD simulations. Here 0 stands for a central Ge atom in a tetrahedron and the numbers 1,2,3,4,5 and 6 represent the neighbors (i.e. 102 indicates the angle between the first and 2nd neighbor of the central Ge atom labeled 0. d) Computed bond angle distributions Se-Ge-Se with increasing pressure for $Ge_{20}Se_{80}$.

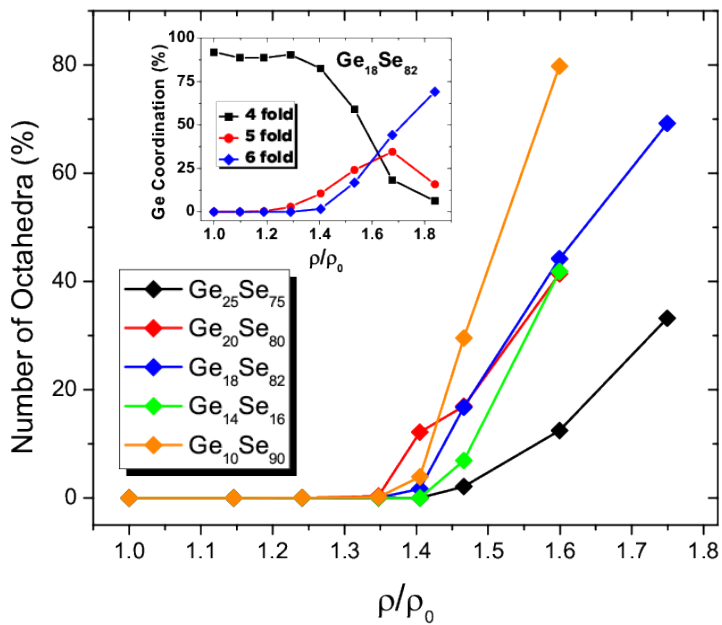

Figure 7: Evolution of 6 fold coordinated Ge atoms as a function of reduced density. The inset shows the changes of various Ge environment in $Ge_{18}Se_{82}$ over the whole pressure range of the AIMD simulations.

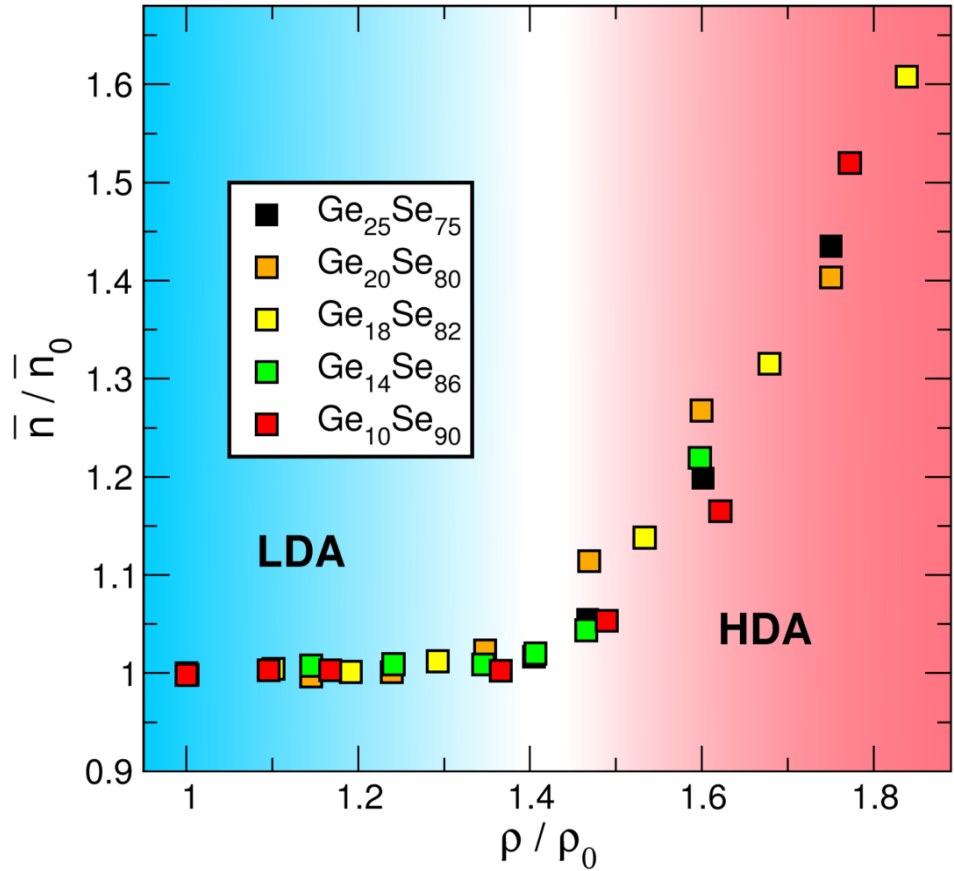

Figure 8: Normalized total coordination number as a function of the reduced density for the selected compositions of $Ge_xSe_{100-x}$ obtained from CPMD calculations. Here the y axis represents the total coordination number at a given $\rho$ divided by the total coordination number at ambient pressure. A global behavior is clearly observed for all the compositions with a marked increase in the coordination number starting around $\rho=\rho_0 = 1.4$.